# An Optical Method for Measuring Local Convective Heat Transfer Coefficients Over 100 W/(m$^2$·K) with Sub-Millimeter Resolution

Tao Chen[1], Puqing Jiang[1,*]

[1]*School of Power and Energy Engineering, Huazhong University of Science and Technology, Wuhan, Hubei 430074, China*

**ABSTRACT**: Conventional methods for measuring the local convective heat transfer coefficient ($h_c$) often rely on simplifying assumptions that can compromise accuracy. Pump-probe methods like time-domain thermoreflectance (TDTR) avoid these assumptions but are limited to $h_c$ values larger than $30\ \mathrm{kW/(m^2 \cdot K)}$ due to modulation frequency constraints. This study introduces an optical-based Square-Pulsed Source (SPS) method, capable of measuring $h_c$ values above $100\ \mathrm{W/(m^2 \cdot K)}$ with sub-millimeter spatial resolution and less than 10% uncertainty. The method was validated through experiments on an air impinging jet, where local Nusselt number distributions were measured and compared to established correlations, demonstrating excellent agreement and revealing detailed convective heat transfer dynamics. This work establishes the SPS method as a new tool for high-precision, localized convective heat transfer analysis, enabling improved design and optimization of thermal systems across diverse applications.

## 1. Introduction

As technology advances, the need for precise thermal management becomes increasingly critical, necessitating measurements of the local convective heat transfer coefficient ($h_c$) with sub-millimeter-scale spatial resolution. One primary motivation is the rising cooling demand for modern electronic devices, where effective cooling solutions, such as microchannel heat sinks and microjet impingement, require high-resolution $h_c$ measurements for optimized design and performance [1]. Beyond electronics, accurate local $h_c$ measurement is crucial in various industrial and scientific applications, such as cooling turbine blades in aerospace engineering [2], targeted thermal therapies in biomedical fields [3], and controlling material properties in advanced manufacturing [4]. Precise $h_c$ measurement not only optimizes convective heat transfer processes and improves energy efficiency but also validates computational models, thereby advancing thermal science and technology.

Currently, the most common method for determining local $h_c$ is the direct method [5]. This approach involves applying a uniform heat flux to the wall surface and simultaneously measuring the local wall and fluid temperatures to calculate $h_c$ based on Newton's law of cooling. However, this method is susceptible to inaccuracies due to several challenges. For instance, unavoidable heat loss complicates the accurate determination of the heat flux transferred to the fluid. Additionally, inserting temperature sensors can disturb the flow field, especially in confined spaces such as microchannels. When the wall temperature is non-uniform, heat conduction within the wall occurs, which is often neglected by assuming one-dimensional heat transfer in the analysis.

While variations of the direct method, such as employing contactless techniques like infrared thermography for wall temperature measurement [6, 7], can address some of these issues, significant limitations remain. The direct method typically requires a large temperature difference between the wall and fluid to minimize measurement uncertainty. However, this large temperature difference causes significant fluid temperature variations, necessitating the use of the "film temperature" concept. Moreover, the direct method struggles with measuring high $h_c$ values, as maintaining the required temperature difference demands a large heat flux, exacerbating the heat loss effect.

Alternative methods, such as the mass transfer analogy [8, 9] and laser interferometry [10, 11], can measure local $h_c$ in specific situations, but each comes with its own set of limitations. The heat-mass analogy method struggles to ensure identical boundary conditions, such as the constant heat flux boundary condition in heat transfer problems, which is challenging to impose in mass transfer problems. The zero wall velocity condition may also be violated due to the presence of a normal velocity component near the wall caused by mass transfer. The laser interferometry method, on the other hand, faces difficulties when measuring in confined spaces.

Optical pump-probe methods offer innovative approaches for measuring local $h_c$. Techniques such as time-domain thermoreflectance (TDTR) and frequency-domain thermoreflectance (FDTR) have been used to measure local convective heat transfer [12-16]. In these methods, a transparent wall coated with a thin metal film undergoes the convective heat transfer process. A modulated pump laser passes through the transparent wall and focuses on the metal film to apply a localized heat flux, while a probe laser detects the temperature response of the metal film at the same location. By developing a thermal model and optimally fitting the measurement signals, the local $h_c$ from the wall surface to the fluid can be extracted. However, TDTR's modulation frequency is limited to a short range from 0.1 to 10 MHz, restricting measurements to $h_c$ values larger than 30 kW/(m²·K) [16].

Recently, a new pump-probe method called Square-Pulsed Source (SPS) has been developed to measure the thermal properties of bulk and thin-film materials [17, 18]. This method allows for the observation of signal variation in the time domain and a wide range of variable modulation frequencies from 1 Hz to 10 MHz. This capability enables the measurement of in-plane thermal diffusivity and cross-plane thermal effusivity across a broad range of materials with thermal conductivities from 0.2 to 2000 W/(m·K) [17], as well as the determination of the full anisotropic thermal conductivity tensor [18]. In this work, we demonstrate that the SPS method can also measure local $h_c$ values greater than 100 W/(m²·K) with a typical uncertainty of less than 10%.

This paper details the basic principles of the SPS method for measuring local $h_c$, applies it to measure the local convective heat transfer coefficient of an air impinging jet, and compares the results with established correlations from the literature. Our findings indicate that the SPS method is a valuable new tool for local convective heat transfer measurement, offering significant advantages in improving measurement accuracy and expanding the measurement range.

**Nomenclature**

| | |
|---|---|
| $a$ | Undetermined complex constant defined in Eq.(3) |
| $b$ | Undetermined complex constant defined in Eq.(3) |
| $A_0$ | the pump power absorbed by the sample, W |
| $A_{\mathrm{norm}}$ | Normalized amplitude signal |
| $C$ | Volumetric heat capacity, $\mathrm{J/(m^3 \cdot K)}$ |
| $D$ | Nozzle diameter, m |
| $f_0$ | Modulation frequency of the pump, Hz |
| $G$ | Interfacial thermal conductance, $\mathrm{W/(m^2 \cdot K)}$ |
| $\hat{G}$ | Green's function |
| $h$ | Thickness, m |
| $h_c$ | Local convective heat transfer coefficient, $\mathrm{W/(m^2 \cdot K)}$ |
| $\bar{h}_c$ | Average convective heat transfer coefficient, $\mathrm{W/(m^2 \cdot K)}$ |
| $H$ | Distance from the nozzle to the wall, m |
| $k$ | Thermal conductivity, $\mathrm{W/(m \cdot K)}$ |
| Nu | Nusselt number |
| $\overline{\mathrm{Nu}}$ | Average Nusselt number |
| Pr | Prandtl number |
| $Q$ | Fourier-Hankel transform of heat flux |
| $Q_{\mathrm{f}}$ | Air flow rate, $\mathrm{m^3/s}$ |
| $r$ | Radial coordinate in cylindrical coordinates, m |
| $r_0$ | RMS average of pump and probe spot radii, m |
| $r_1$ | $1/e^2$ radius of the pump laser spot, m |
| $r_2$ | $1/e^2$ radius of the probe laser spot, m |
| Re | Reynolds number |
| $S_\xi$ | Sensitivity coefficient of $A_{\mathrm{norm}}$ to $\xi$ |
| $t$ | Time, s |
| $T$ | Temperature, K |
| $T_0$ | Ambient temperature, K |
| $u$ | Air jet velocity, $\mathrm{m/s}$ |
| $z$ | Axial coordinate in cylindrical coordinates, m |

*Greek symbols*

| | |
|---|---|
| $\alpha$ | Thermal diffusivity, $\mathrm{m^2/s}$ |
| $\gamma$ | A complex number defined in Eq.(4) |
| $\theta$ | Excess temperature, K |
| $\Theta$ | Fourier−Hankel transform of $\theta$ |
| $\lambda$ | A complex number defined in Eq.(2) |
| $\nu$ | Kinematic viscosity, $\mathrm{m^2/s}$ |
| $\xi$ | Any parameter in the heat transfer model |
| $\rho$ | Variable for Hankel transform |
| $\omega$ | Angular frequency |

*Subscripts*

| | |
|---|---|
| m | Physical properties of the metal film |
| sub | Physical properties of the substrate |
| $n$ | The $n$-th layer, $n = 0, 1$ |
| f | Physical properties of dry air |

*Superscripts*

| | |
|---|---|
| $+$ | Below the heat source |
| $-$ | Above the heat source |

## 2. Experimental Method

### *2.1 Square-Pulsed Source Method*

The principle of the SPS method has been described in previous studies [17, 18]. In brief, a thin metal transducer layer is deposited on the sample surface to absorb heat from a square-wave-modulated pump laser. The temperature response of the sample is then detected via thermoreflectance by monitoring changes in the transducer's reflectance, which are proportional to the temperature variation as long as the temperature change remains below approximately 10 K [19]. The reflected probe beam is collected by a balanced amplified detector, which converts the optical signal into an electrical output sent to a periodic waveform analyzer (PWA), yielding the amplitude of the temperature change over a single square-wave heating cycle. These measured signals are then normalized for both the amplitude and time and compared to predictions from a thermal model. The model's fitting parameters are iteratively adjusted until the simulated signals best match the measured data, enabling the extraction of parameters like $h_c$.

Figure 1 shows a schematic diagram of the SPS system used for local $h_c$ measurements. The sample being measured is a transparent, low-conductivity substrate coated with a thin metal transducer film. A laser beam passes through the transparent substrate from the opposite side, striking the metal film at the metal-substrate interface. Convective heat transfer at the metal film surface serves as a boundary condition for the heat diffusion process within the sample.

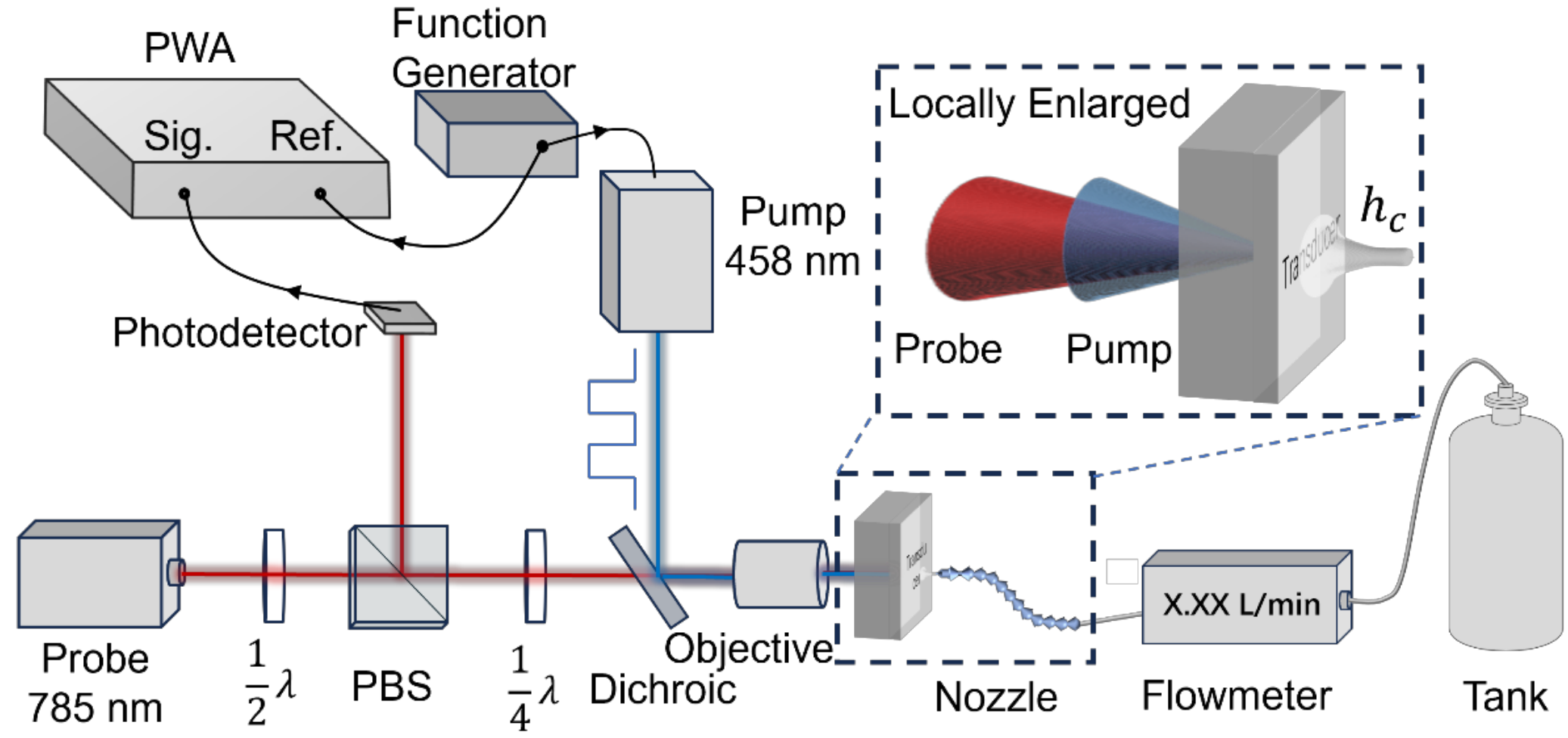

**Fig. 1.** Schematic diagram of the experimental system for measuring the local convective heat transfer coefficient of an air jet using the Square-Pulsed Source (SPS) method.

*2.2 Thermal Model*

The thermal model serves to generate simulated signals and best fit the measured data, enabling inverse extraction of the target thermal properties. Figure 2 shows a schematic of the two-layered wall structure used in convective heat transfer experiments. The heat flux applied at the interface between the two layers diffuses in both directions, with the temperature response detected at the same location. The thermal model is based on the heat diffusion processes in both layers, with the convective heat transfer at the surface of the transducer film treated as a boundary condition.

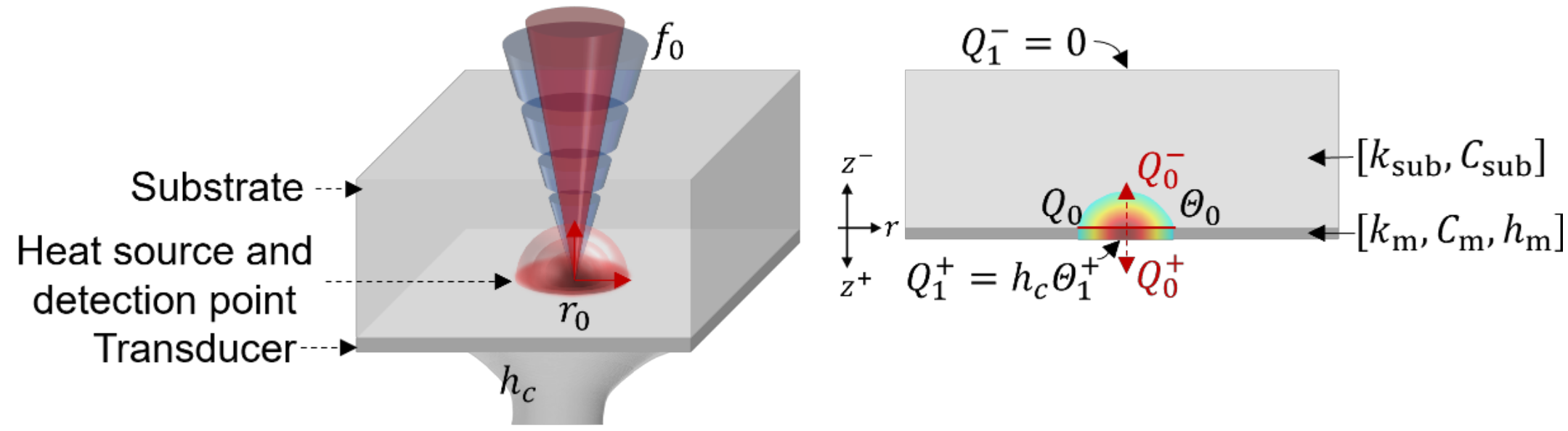

**Fig. 2.** Schematic of the two-layered wall structure used in convective heat transfer experiment: A modulated pump beam heats the interface between the two layers, while the temperature response at the same location is monitored by a probe beam. The applied heat flux propagates into both layers in opposite directions, with the substrate assumed to be infinitely thick. The surface of the transducer film undergoes convective heat transfer, while the substrate remains insulated.

The governing equation for heat diffusion in each layer of the wall, expressed in cylindrical coordinates, is given by: [20]

$$\frac{1}{\alpha}\frac{\partial \theta}{\partial t} = \frac{1}{r}\frac{\partial}{\partial r}\left(r\frac{\partial \theta}{\partial r}\right) + \frac{\partial^2 \theta}{\partial z^2} \tag{1}$$

Here, $\theta = T - T_0$ represents the excess temperature (with $T_0$ as the ambient temperature), and $\alpha = k/C$ is the thermal diffusivity, with $k$ being the thermal conductivity and $C$ the volumetric heat capacity of the material. The variables $t$, $r$, and $z$ represent time, radial, and axial coordinates, respectively.

This partial differential equation can be transformed into an ordinary differential equation using Fourier and Hankel transforms [17], yielding:

$$\frac{\partial^2 \Theta}{\partial z^2} = \lambda^2 \Theta \tag{2}$$

Here, $\Theta$ is the frequency-domain representation of $\theta$, and $\lambda^2 = 4\pi^2\rho^2 + i\omega/\alpha$, with $\rho$ as the Hankel transform variable and $\omega$ as the angular frequency in the Fourier transform.

The general solution of Eq. (2) is:

$$\Theta = ae^{\lambda z} + be^{-\lambda z} \tag{3}$$

where $a$ and $b$ are undetermined complex constants that depend on the material properties and boundary conditions.

Using the temperature expression from Eq. (3) and applying Fourier's law of heat conduction, the heat flux in the frequency domain is obtained as:

$$Q = -k\frac{d\Theta}{dz} = \gamma(-ae^{\lambda z} + be^{-\lambda z}) \tag{4}$$

where $\gamma = k\lambda$.

In the thermal system shown in Fig. 2, the heat flux supplied at the substrate-film interface is denoted as $Q_0$, which is divided into two components: $Q_0^+$, the heat flux diffusing through the metal film and convected away, and $Q_0^-$, the heat flux diffusing into the semi-infinite substrate. Therefore, we have:

$$Q_0 = Q_0^+ + Q_0^- \tag{5}$$

At the interface ($z = 0$), the temperature and heat flux can be obtained from Eqs. (3) and (4) as:

$$\Theta_0 = a_{\mathrm{m}} + b_{\mathrm{m}} = a_{\mathrm{sub}} + b_{\mathrm{sub}} \tag{6}$$

$$Q_0^+ = \gamma_{\mathrm{m}}(-a_{\mathrm{m}} + b_{\mathrm{m}}) \tag{7}$$

$$Q_0^- = \gamma_{\mathrm{sub}}(-a_{\mathrm{sub}} + b_{\mathrm{sub}}) \tag{8}$$

Here, the subscripts "m" and "sub" represent the properties for the metal film and the substrate, respectively.

At the surface of the metal film, the convective heat transfer boundary condition is:

$$Q_1^+ = h_c\Theta_1^+ \tag{9}$$

The surface temperature $\Theta_1^+$ and heat flux $Q_1^+$ of the metal film can be obtained from Eqs. (3) and (4) by setting $z = h_{\mathrm{m}}$ (the thickness of the metal film):

$$\Theta_1^+ = a_{\mathrm{m}}e^{\lambda_{\mathrm{m}}h_{\mathrm{m}}} + b_{\mathrm{m}}e^{-\lambda_{\mathrm{m}}h_{\mathrm{m}}} \tag{10}$$

$$Q_1^+ = \gamma_{\mathrm{m}}\left(-a_{\mathrm{m}}e^{\lambda_{\mathrm{m}}h_{\mathrm{m}}} + b_{\mathrm{m}}e^{-\lambda_{\mathrm{m}}h_{\mathrm{m}}}\right) \tag{11}$$

On the back side of the substrate, the heat flux $Q_1^-$ is zero because the substrate is modeled as semi-infinite, which gives:

$$a_{\mathrm{sub}} = 0 \tag{12}$$

Combining Eqs. (5) to (12), we can obtain the Green's function in the frequency

domain as:

$$\hat{G}(\rho,\omega)=\frac{\Theta_0}{Q_0}=\frac{1}{\gamma_{\mathrm{sub}}+\gamma_{\mathrm{m}}\frac{b_{\mathrm{m}}-a_{\mathrm{m}}}{b_{\mathrm{m}}+a_{\mathrm{m}}}}=\frac{1}{\gamma_{\mathrm{sub}}+\gamma_{\mathrm{m}}\frac{\gamma_{\mathrm{m}}\cdot\tanh(\frac{h_{\mathrm{m}}}{k_{z\mathrm{m}}}\gamma_{\mathrm{m}})+h_c}{h_c\cdot\tanh(\frac{h_{\mathrm{m}}}{k_{z\mathrm{m}}}\gamma_{\mathrm{m}})+\gamma_{\mathrm{m}}}} \tag{13}$$

In SPS measurements, the heat flux imposed by the pump laser is spatially distributed as a Gaussian with a radius of $r_1$ and temporally as a square wave function with a frequency of $f_0$. After performing the Hankel transform and Fourier transform, the expression for $Q_0$ is:

$$Q_0(\rho,\omega)=A_0\exp\left(-\frac{\pi^2\rho^2r_1^2}{2}\right)\left(\frac{\delta(\omega)}{2}+\frac{1}{\pi}\sum_{n=1}^{\infty}i\,\frac{(\delta(\omega+(2n-1)2\pi f_0)-\delta(\omega-(2n-1)2\pi f_0))}{2n-1}\right) \tag{14}$$

where $A_0$ is the power of the pump laser absorbed by the metal transducer.

Substituting the expression for $Q_0$ into Eq. (13) gives the temperature response $\Theta_0$ in the frequency domain. Applying the inverse Hankel and Fourier transforms to $\Theta_0$ yields the temperature expression $\theta(r,t)$ in the time domain. The spatially weighted average of $\theta(r,t)$ by a probe laser with a radius of $r_2$ is then calculated, resulting in the signal variation over time detected by the photodetector as:

$$\Delta T(t)=\frac{A_0}{2}\int_{-\infty}^{\infty}\hat{G}(\rho,0)\exp(-\pi^2\rho^2r_0^2)\,2\pi\rho\mathrm{d}\rho$$
$$-2A_0\mathrm{Re}\left(\sum_{n=1}^{\infty}\frac{i}{(2n-1)\pi}e^{i(2n-1)\omega_0t}\right)\int_{-\infty}^{\infty}\hat{G}(\rho,(2n-1)\omega_0)\exp(-\pi^2\rho^2r_0^2)\,2\pi\rho\mathrm{d}\rho \tag{15}$$

where $r_0=\sqrt{\frac{r_1^2+r_2^2}{2}}$, and $\mathrm{Re}(x)$ means taking the real part of the complex number $x$.

The solution for $\Delta T(t)$ must be obtained numerically. By normalizing this numerical solution, it can be directly compared with the measured signals to extract the parameters of interest.

For simplicity, the model derived above neglects the interfacial thermal conductance ($G$) between the metal film and the substrate. This assumption is justified because $G$, which is on the order of $100\ \mathrm{MW/(m^2\cdot K)}$, is several orders of magnitude higher than $h_c$, and thus has a negligible effect on the heat diffusion process. Additionally, the optical penetration of the laser beam into the metal film is ignored, and a surface heat flux is assumed instead. This is reasonable, as the optical penetration depth of the laser beam (~30 nm) is much smaller than the laser spot diameter (~100 μm). The insulated boundary condition on the backside of the substrate is also valid, as the substrate thickness (2 mm) is significantly larger than the size of the heated region (~0.1 mm), ensuring that the thermal wave within the substrate does not reach the backside. No further assumptions are made in the thermal model, ensuring its accuracy in describing the entire heat transfer process during the convective heat transfer experiments.

As shown in Figure 2, the thermal model involves eight parameters: the modulation frequency $f_0$, laser spot radius $r_0$, thermal conductivity $k_{\mathrm{m}}$, volumetric heat capacity $C_m$, and thickness $h_{\mathrm{m}}$ of the metal film; thermal conductivity $k_{\mathrm{sub}}$ and volumetric heat capacity $C_{\mathrm{sub}}$ of the substrate; and the convective heat transfer coefficient $h_c$. Among these, $f_0$ is a set value, while parameters such as $r_0$, $k_{\mathrm{m}}$, and $h_{\mathrm{m}}$ can be independently calibrated. The value of $C_{\mathrm{m}}$ can be obtained from the literature. The thermal properties $k_{\mathrm{sub}}$ and $C_{\mathrm{sub}}$ of the substrate can either be sourced from the literature or measured independently using the SPS method at high modulation frequencies ranging from 100 Hz to 10 MHz [17]. This leaves $h_c$ as the only parameter to be fitted.

A flowchart illustrating the process of extracting $h_c$ from the SPS experiment is shown in Figure 3. First, SPS measurements are conducted at a low frequency of approximately 1 Hz to acquire the experimental signals, which are then normalized in both amplitude and time. The next step involves inputting known parameters along with an initial guess for $h_c$ into the thermal model to generate simulated signals. The value of $h_c$ is then iteratively adjusted until the simulated signals closely match the experimental data.

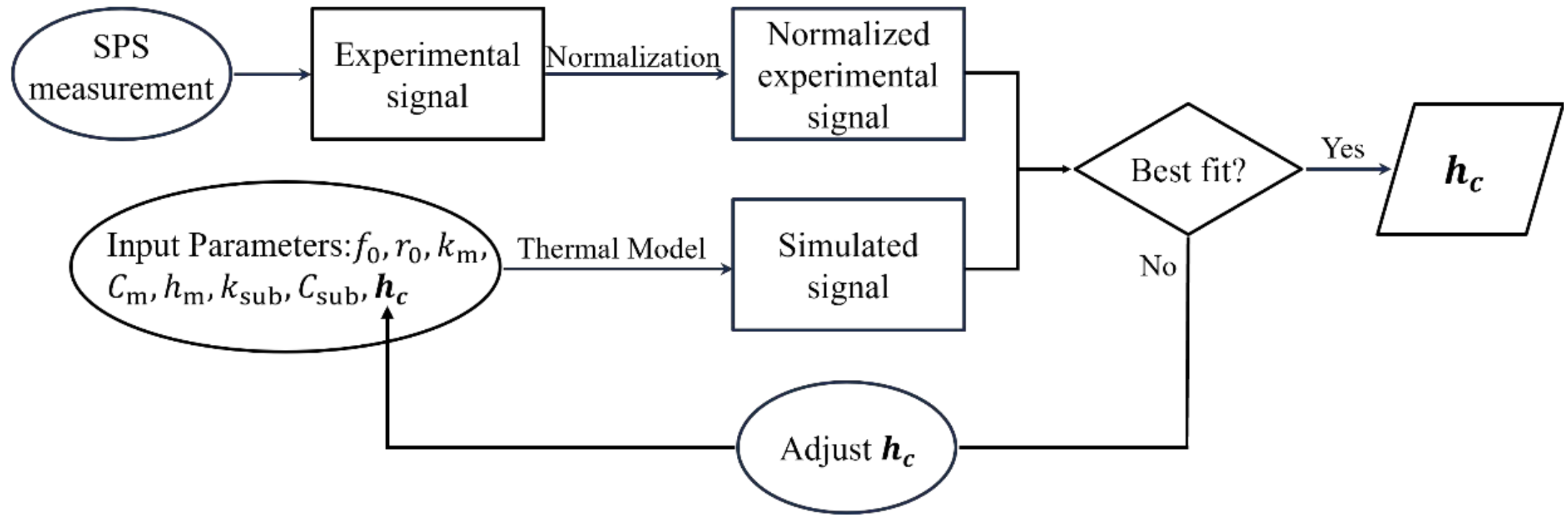

**Fig. 3.** The flowchart for extracting $h_c$ from SPS experiments.

*2.3 Sensitivity Analysis*

The effect of various parameters on the normalized amplitude signal is described by the sensitivity coefficient $S_\xi$ [21]:

$$S_\xi = \frac{\partial A_{\text{norm}}/A_{\text{norm}}}{\partial \xi/\xi} \tag{16}$$

where $A_{\text{norm}}$ is the normalized amplitude signal, and $\xi$ represents any parameter to be analyzed. According to this definition, $S_\xi$ indicates that a 1% change in parameter $\xi$ will cause a $S_\xi$% change in the signal $A_{\text{norm}}$.

We note that alternative definitions for the sensitivity coefficient exist in the literature, such as $\frac{\partial A_{\text{norm}}}{\partial \xi/\xi}$ and $\frac{\partial A_{\text{norm}}}{\partial \xi}$ [21]. However, Eq. (16) is more suitable for the SPS experiments as it better captures the degree of impact that parameter has on the signal. In Eq. (16), the percentage change in the signal $A_{\text{norm}}$ is evaluated because the signal spans several orders of magnitude, ranging from 0.001 to 1 in SPS experiments. This necessitates plotting on a logarithmic scale to highlight low-amplitude signals. Similarly, we evaluate the percentage change in the parameter $\xi$ to enable comparisons across multiple parameters with varying units and orders of magnitudes.

Conducting a sensitivity analysis prior to measurements is essential for optimizing experimental design and selecting the most suitable measurement conditions to achieve accurate and reliable results. Through detailed sensitivity analysis, we determined that accurately measuring convective heat transfer coefficients as low as 100 W/(m²·K) requires two key design choices: using low thermal conductivity materials for the transparent substrate and employing a low modulation frequency (~1 Hz). These adjustments minimize the relative contribution of the term $\gamma_{\mathrm{sub}}$ in the denominator of the Green's function $\hat{G}(\rho,\omega)$ (Eq. (13)), thereby amplifying the influence of the term $\gamma_{\mathrm{m}} \frac{h_c + \tanh(\frac{h_{\mathrm{m}}}{k_{z\mathrm{m}}}\gamma_{\mathrm{m}})\cdot\gamma_{\mathrm{m}}}{h_c\cdot\tanh(\frac{h_{\mathrm{m}}}{k_{z\mathrm{m}}}\gamma_{\mathrm{m}}) + \gamma_{\mathrm{m}}}$ that contains $h_c$. This design enhances the signal's sensitivity to $h_c$. However, the contribution of $\gamma_{\mathrm{sub}}$ remains non-negligible and must still be considered in the analysis. For higher convective heat transfer coefficients, such as those on the order of 1 kW/(m²·K), these conditions can be relaxed; for instance, the modulation frequency can be increased to 10 Hz. These insights provide practical guidance for optimizing experimental setups and ensuring accurate measurements of a wide range of convective heat transfer coefficients.

*2.4 Uncertainty Analysis*

The uncertainty in the best-fit value of $h_c$ obtained from SPS experiments can be estimated using an error propagation formula:

$$\frac{\delta_{h_c}}{h_c} = \sqrt{\sum_{\xi}\left(\frac{S_{\xi}}{S_{h_c}}\frac{\delta\xi}{\xi}\right)^2} \tag{17}$$

where $\xi$ represents all controlled parameters. This formula assumes independent uncertainties of each parameter. Moreover, it does not consider the quality of the fitting process and may result in an uncertainty that varies with the *x*-coordinate of the signals, complicating interpretation.

To address these limitations, we employed a more advanced approach for estimating the uncertainty in $h_c$, based on a comprehensive error propagation formula. This formula is derived from the best fit of a set of signals using the least-squares regression method and has been validated against Monte Carlo simulations [22, 23]. In addition to propagating errors from input parameters, it also incorporates uncertainties from experimental noise and the quality of the fitting process. This approach was first introduced by Yang and Schmidt [22] to analyze uncertainties in FDTR experiments, and has since been updated and widely adopted in other thermoreflectance studies [24-27]. The details of this error propagation method are provided below.

The best-fitting process involves minimizing the difference between the experimental signals and the corresponding model predictions. This is mathematically represented as:

$$J = \sum_{i=1}^{N}\left(\frac{g(h_c,\boldsymbol{X_P},t_i)}{y(t_i)} - 1\right)^2 \tag{18}$$

Here, $y(t_i)$ is the measured signal at the $i$-th time point $t_i$, and $g(h_c, \boldsymbol{X_P}, t_i)$ represents the simulated signal at the same time point, based on inputs of the fitting parameter $h_c$ and the controlled parameter vector $\boldsymbol{X_P}$. In the current case, $\boldsymbol{X_P} = (r_0, k_{\mathrm{m}}, C_{\mathrm{m}}, h_{\mathrm{m}}, k_{\mathrm{sub}}, C_{\mathrm{sub}})^T$. $N$ is the total number of data points.

The gradient of $J$ must be zero at the best-fit value of $h_c$:

$$\sum_{i=1}^{N}\frac{2\left(g(\hat{h}_c,\widehat{\boldsymbol{X}}_{\boldsymbol{P}},t_i)-y(t_i)\right)}{y^2(t_i)}\left(\frac{\partial g}{\partial h_c}\right)_{\hat{h}_c} = 0, \tag{19}$$

where $\widehat{\boldsymbol{X}}_{\boldsymbol{P}}$ represents the distributions of all possible $\boldsymbol{X_P}$, and $\hat{h}_c$ is the distribution of the correspondingly best-fitted $h_c$. Let $\boldsymbol{X_P^0}$ and $h_c^0$ represent their mean values. The function $g\left(\hat{h}_c, \widehat{\boldsymbol{X}}_{\boldsymbol{P}}, t_i\right)$ can be approximated using a first-order Taylor expansion around $(h_c^0, \boldsymbol{X_P^0})$:

$$g\left(\hat{h}_c,\widehat{\boldsymbol{X}}_{\boldsymbol{P}},t_i\right) \approx g\left(h_c^0,\boldsymbol{X_P^0},t_i\right) + \left(\frac{\partial g}{\partial h_c}\right)_{h_c^0,X_P^0}\left(\hat{h}_c - h_c^0\right) + \sum_{j=1}^{q}\left(\frac{\partial g}{\partial p_j}\right)_{h_c^0,X_P^0}\left(\hat{p}_j - p_j^0\right) \tag{20}$$

where $p_j$ is the $j$-th element of the vector $\boldsymbol{X_P}$, and $q$ is the length of $\boldsymbol{X_P}$.

Substituting Eq. (20) into Eq. (19) and ignoring higher-order terms, we get:

$$\sum_{i=1}^{N}\frac{1}{y^2(t_i)}\left[g\left(h_c^0,\boldsymbol{X_P^0},t_i\right) - y(t_i) + \left(\frac{\partial g}{\partial h_c}\right)_{h_c^0,X_P^0}\left(\hat{h}_c - h_c^0\right) + \sum_{j=1}^{q}\left(\frac{\partial g}{\partial p_j}\right)_{h_c^0,X_P^0}\left(\hat{p}_j - p_j^0\right)\right]\left(\frac{\partial g}{\partial h_c}\right)_{h_c^0,X_P^0} = 0 \tag{21}$$

In matrix form, this simplifies to:

$$\boldsymbol{J_C^T D}(\boldsymbol{E} - \boldsymbol{F}) - \boldsymbol{J_C^T D J_P}\left(\widehat{\boldsymbol{X}}_{\boldsymbol{P}} - \boldsymbol{X_P^0}\right) = \boldsymbol{J_C^T D J_C}\left(\hat{h}_c - h_c^0\right) \tag{22}$$

where $\boldsymbol{D} = \mathrm{diag}(\frac{1}{y^2(t_1)}, \frac{1}{y^2(t_2)}, \cdots, \frac{1}{y^2(t_N)})$, $\boldsymbol{E}$ is the column vector of measured signals, and $\boldsymbol{F}$ is the vector of model-predicted signals at $(h_c^0, \boldsymbol{X_P^0})$. The matrices $\boldsymbol{J_C}$ and $\boldsymbol{J_P}$ are given by:

$$\boldsymbol{J_C} = \begin{pmatrix} \frac{\partial g(t_1)}{\partial h_c}|_{h_c^0, \boldsymbol{X_P^0}} \\ \vdots \\ \frac{\partial g(t_N)}{\partial h_c}|_{h_c^0, \boldsymbol{X_P^0}} \end{pmatrix} \tag{23}$$

and

$$\boldsymbol{J_P} = \begin{pmatrix} \frac{\partial g(t_1)}{\partial p_1}|_{h_c^0, \boldsymbol{X_P^0}} & \cdots & \frac{\partial g(t_1)}{\partial p_q}|_{h_c^0, \boldsymbol{X_P^0}} \\ \vdots & \ddots & \vdots \\ \frac{\partial g(t_N)}{\partial p_1}|_{h_c^0, \boldsymbol{X_P^0}} & \cdots & \frac{\partial g(t_N)}{\partial p_q}|_{h_c^0, \boldsymbol{X_P^0}} \end{pmatrix} \tag{24}$$

Defining $\boldsymbol{\Sigma_{CC}} = \boldsymbol{J_C^T D J_C}$ and $\boldsymbol{\Sigma_{CP}} = \boldsymbol{J_C^T D J_P}$, and assuming $\boldsymbol{\Sigma_{CC}}$ is non-singular, we can express $\hat{h}_c$ as:

$$\hat{h}_c = \boldsymbol{\Sigma_{CC}^{-1} J_C^T D}(\boldsymbol{E} - \boldsymbol{F}) - \boldsymbol{\Sigma_{CC}^{-1} \Sigma_{CP}}(\boldsymbol{\widehat{X}_P} - \boldsymbol{X_P^0}) + h_c^0 \tag{25}$$

The variance of $\hat{h}_c$, representing the uncertainty in $h_c$, is given by:

$$\mathrm{Var}[\hat{h}_c] = \sigma_{h_c}^2 = \boldsymbol{\Sigma_{CC}^{-1} J_C^T D}\mathrm{Var}[\boldsymbol{E} - \boldsymbol{F}]\boldsymbol{D}^T \boldsymbol{J_C \Sigma_{CC}^{-1}} + \boldsymbol{\Sigma_{CC}^{-1} \Sigma_{CP}}\mathrm{Var}[\boldsymbol{\widehat{X}_P}]\boldsymbol{\Sigma_{CP}^T \Sigma_{CC}^{-1}} \tag{26}$$

where $\mathrm{Var}[\boldsymbol{E} - \boldsymbol{F}]$ is an $N \times N$ diagonal matrix with elements $\left(y(t_i) - g(h_c^0, X_P^0, t_i)\right)^2$, and $\mathrm{Var}[\boldsymbol{\widehat{X}_P}]$ is a $q \times q$ diagonal matrix with elements $\sigma_{p_j}^2$, where $2\sigma_{p_j}$ is the uncertainty of the controlled parameter $p_j$. Equation (26) comprises two terms: the first accounts for uncertainty from experimental noise and fitting quality, while the second reflects uncertainty propagated from errors in the controlled variables. The final uncertainty in $h_c$ is $2\sigma_{h_c}$.

## 3. Measurement Results and Heat Transfer Analysis of an Air Impinging Jet

This study demonstrates the feasibility and accuracy of the SPS method for measuring the local convective heat transfer coefficient $h_c$ by analyzing the heat transfer characteristics of an air impinging jet. Impinging jet cooling is widely regarded as an efficient means to enhance the convective heat transfer rate between a fluid and a surface. In this process, the jet directly strikes the solid surface, where intense turbulent mixing and boundary layer renewal at the point of impact significantly enhance the local heat transfer coefficient [28]. However, the heat transfer intensity within an impinging jet is inherently non-uniform. Precise measurement and a deeper understanding of these thermal dynamics are essential for optimizing impingement cooling systems. Such insights allow engineers to identify areas with the highest cooling efficiency and to design nozzle arrangements and operating conditions that maximize heat transfer at targeted locations.

In the experiments conducted in this study, the wall material was optical-grade acrylic (PMMA), coated with a thin aluminum (Al) film deposited via magnetron sputtering to serve as the transducer layer. At room temperature, PMMA has a thermal conductivity of $0.19\ \mathrm{W/(m\cdot K)}$ and a heat capacity of $1.644\ \mathrm{MJ/(m^3\cdot K)}$ [29]. Its low thermal conductivity facilitates efficient heat transfer toward convection, making it favorable for measuring $h_\mathrm{c}$. The Al film, with its high thermoreflectance coefficient at the probe wavelength of 785 nm, acts as an effective transducer for thermoreflectance measurements. The Al film's thickness was measured to be 93 nm using a step profiler. The thermal conductivity $k_\mathrm{m}$ of the Al film was determined from its electrical resistivity using the Wiedemann-Franz law, yielding a value of $45 \pm 5\ \mathrm{W/(m\cdot K)}$, and the resistivity was measured using the van der Pauw method. The Al film's heat capacity was taken from the literature database as $2.42\ \mathrm{MJ/(m^3\cdot K)}$ [30]. The $1/e^2$ radii of the pump and probe laser spots were determined using the knife-edge method.

To conduct the convective heat transfer experiment, compressed air at room temperature was ejected from a 2 mm diameter nozzle ($D = 2$ mm) at a flow rate of $Q_\mathrm{f} = 5\ \mathrm{L/min}$, directed towards the metal film surface positioned 15 mm ($H = 15$ mm) away from the nozzle. This operating condition corresponds to a Reynolds number (Re) of $\mathrm{Re} = uD/\nu = 3520$. A laser spot size of $r_0 = 29\ \mu\mathrm{m}$ and a modulation frequency of $f_0 = 1$ Hz were used for the SPS measurements.

Accurate measurements of the impinging jet require precise alignment of the laser beam, wall plate, and nozzle jet. To achieve this, the PMMA plate was positioned vertically and perpendicular to the laser beam. The nozzle, mounted on an XYZ stage, was first brought close to the plate to verify that the air jet direction was perpendicular to the wall surface. The wall plate was then temporarily removed to allow precise alignment of the laser beam's focus with the center of the nozzle. After alignment, the wall plate was reinstalled, and the nozzle was retracted to a specified distance from the plate before starting the experiment. The relative positions of the nozzle and wall plate were then fixed and mounted together on a bigger XYZ stage. During measurements of the local $h_\mathrm{c}$ at various positions around the jet center, the laser beams remained fixed while both the nozzle and wall plate were moved synchronously to repeat the measurements.

Figure 4 presents an example of the measured signals and data processing when measured at the center of the impinging jet region ($r = 0$). The signals measured over a complete heating cycle are shown in Fig. 4(a), where the symbols represent the experimental data and the curve represents the simulated signals obtained from the thermal model, optimized for the best fit. The heating and cooling segments of the temperature response signals are plotted on logarithmic scales in Fig. 4(b) and (c), respectively, for a more detailed examination of the fit quality between the simulated and measured signals. The simulated signals with ±30% variation in the best-fit $h_c$ are shown as dashed lines in Fig. 4(b) and (c), demonstrating the high sensitivity of the measured signal to $h_c$. The sensitivity coefficients of the measured signals to all the parameters in the thermal system are shown in Fig. 4(d) and (e). The sensitivity curves show that, for both the heating and cooling segments of the temperature response, the signals are primarily sensitive to the combined parameter $k_\mathrm{sub}/(C_\mathrm{sub}r_0^2)$ and the convective heat transfer coefficient $h_c$. Here, $k_\mathrm{sub}/C_\mathrm{sub}$ is the thermal diffusivity of PMMA and can be sourced from the literature, with an uncertainty of around 3%. The laser spot radius $r_0$ can be further calibrated through SPS measurements on the same sample under identical configurations but without air impingement flow. Once these parameters are determined, $h_c$ can be determined by best-fitting the measured signals, yielding a value of $h_c = 700\ \mathrm{W/(m^2 \cdot K)}$ with an estimated error of 7%.

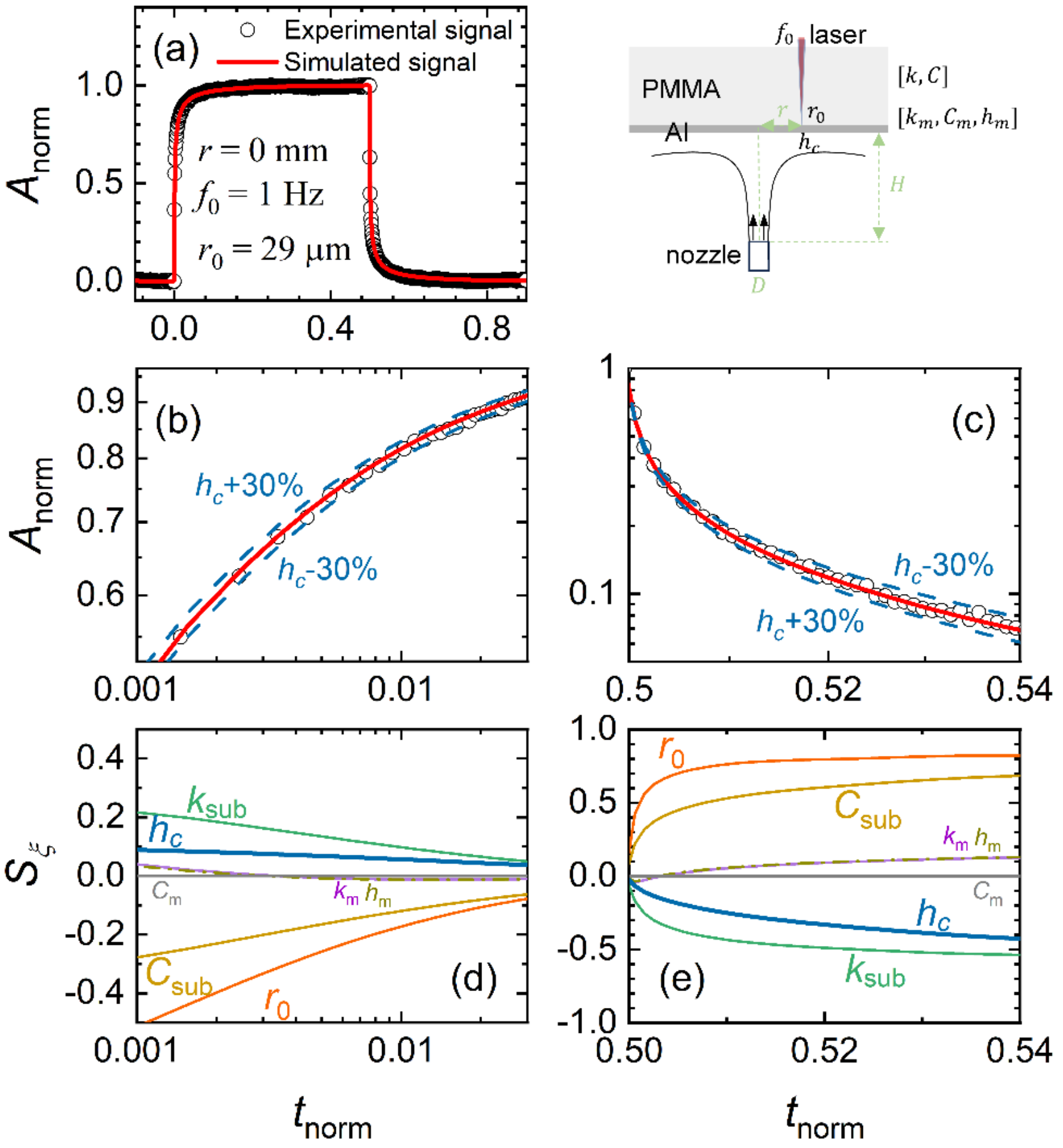

**Fig. 4.** SPS experimental signals and sensitivity curves of the air impinging jet, measured using a modulation frequency of 1 Hz and a spot radius of 29 μm at the center of the impingement region.

The experiments shown in Fig. 4 were repeated, varying the radial distance *r* of the detection point from the impingement center. Figure 5 shows the variation of the local Nusselt number (Nu) with the dimensionless radial distance (*r/D*) at a fixed Re number of 3520, for different dimensionless nozzle-to-wall distances (*H/D*). Here, the Nusselt number is defined as $\mathrm{Nu} = h_c D / k_f$, where $h_c$ is the local convective heat transfer coefficient, *D* is the nozzle diameter, and $k_f$ is the thermal conductivity of dry air, taken as $0.026\ \mathrm{W/(m \cdot K)}$ at room temperature [31].

Figure 5 reveals the complex heat transfer dynamics within the impinging jet convective heat transfer system. Initially, Nu reaches a peak in the stagnation region, where the high-speed jet impingement and intense turbulence generate exceptionally high local heat transfer efficiency. As *r/D* increases, Nu declines, reflecting the jet's deceleration due to an expanding contact area with the surface, reduced turbulence, and diminished kinetic energy. A second peak emerges in the transition region, likely caused by interactions between the jet and the surrounding fluid, such as vortex formation or fluid reattachment to the surface, which momentarily amplifies turbulence intensity. Beyond this region, the jet spreads further, losing momentum and transitioning into a standard wall jet. Consequently, the heat transfer capability decreases, leading to a continued decline in Nu values.

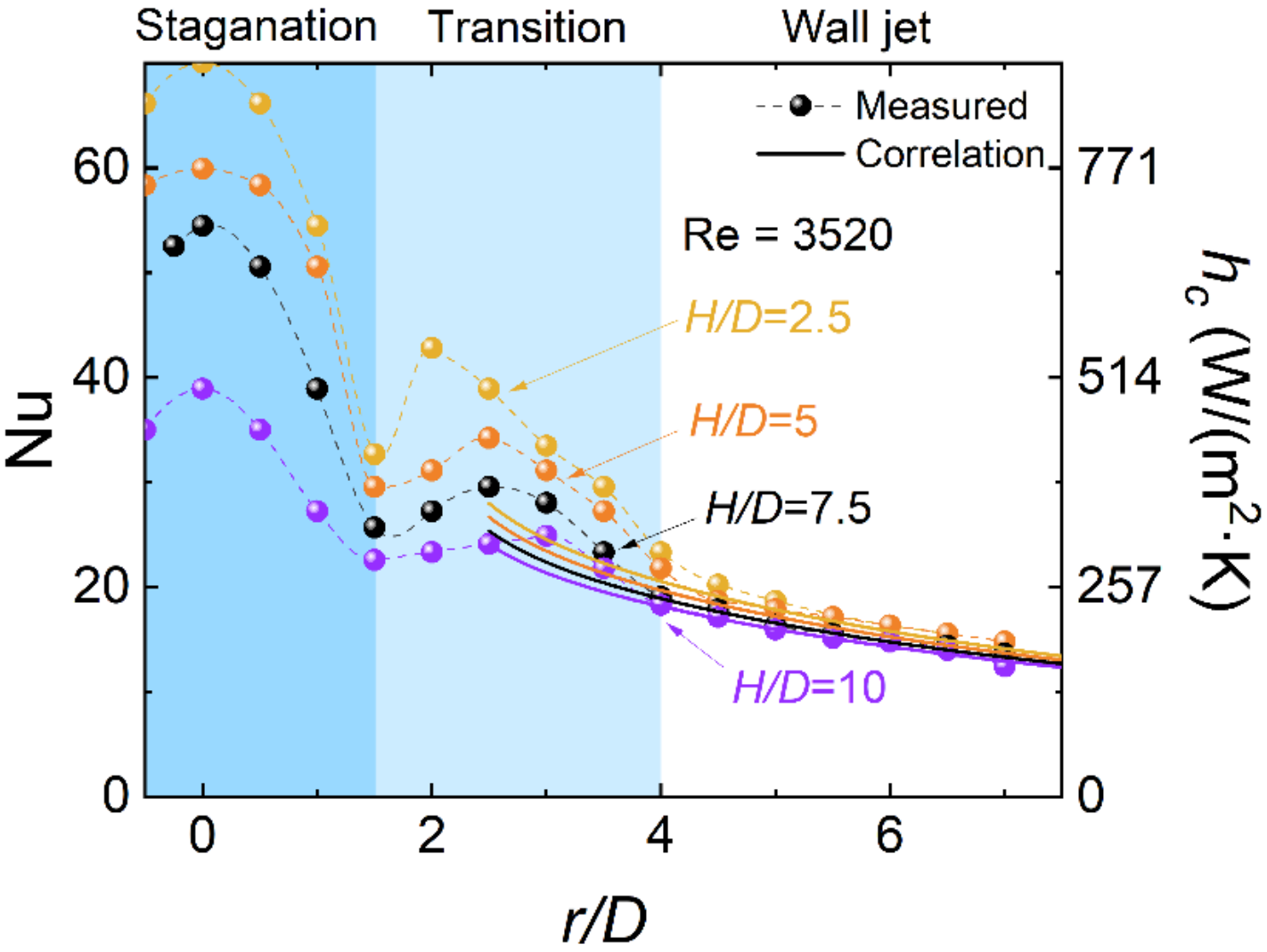

**Fig. 5.** Variation of the local Nusselt number (Nu) with the dimensionless radial distance ($r/D$) at a fixed Reynolds number of 3520 for different dimensionless nozzle-to-wall distances ( $H/D$ ). The symbols connected by dashed lines represent measurements obtained using the SPS method, while the solid curves are predictions based on literature correlations.

Extensive research has been conducted on impinging jet convective heat transfer in the literature. Martin [32] provided a comprehensive review of the subject, covering both single and array nozzle configurations. However, empirical correlations for the average Nusselt number ($\overline{\mathrm{Nu}}$) in the deceleration wall jet region remain the only ones available in the literature [32]. This limitation arises partly due to the complex nature of the stagnation and transition regions. One such empirical correlation is given as:

$$\overline{\mathrm{Nu}} \equiv \bar{h}_c D/k_f = 2\frac{D}{r}\frac{1-1.1D/r}{1+0.1(H/D-6)D/r}\mathrm{Re}^{1/2}(1+0.005\mathrm{Re}^{0.55})^{0.5}\mathrm{Pr}^{0.42} \tag{29}$$

where $\bar{h}_c$ is the average convective heat transfer coefficient from the jet center to a disk of radius $r$. The applicable range of Eq. (29) is: $2.5 \le r/D \le 7.5$, $2 \le H/D \le 12$, and $2\times10^3 \le \mathrm{Re} \le 4\times10^5$.

Using the relationship between the average Nusselt number $\overline{\mathrm{Nu}}$ and the local Nusselt number Nu, i.e., $\overline{\mathrm{Nu}} = \frac{1}{\pi r^2}\int_0^r \mathrm{Nu}(x) 2\pi x \mathrm{d}x$, the local Nusselt number can be derived as:

$$\mathrm{Nu}(r) = \frac{1}{2\pi r}\frac{\mathrm{d}}{\mathrm{d}r}\left(\overline{\mathrm{Nu}}\pi r^2\right) \tag{30}$$

Predictions based on Eq. (30) are shown as solid curves in Fig. 5 and compared with the measurements from this study. In the wall jet region ($r/D \geq 4$), the predictions match the experimental data well, and both show minimum dependence on the dimensionless nozzle-to-wall distance $H/D$. However, in the stagnation and transition regions, significant differences are observed for varying $H/D$ ratios. This indicates that these regions are highly sensitive to turbulence levels at the nozzle outlet. Due to the complex dynamics in these regions, effective empirical correlations are currently unavailable.

The comparisons presented above validate the credibility of our method. This novel optical approach for measuring local convective heat transfer effectively addresses the challenges encountered by convectional direct methods. As an indirect technique, the SPS method extracts $h_c$ by best fitting normalized transient temperature response signals, eliminating the need for absolute temperature or heat flux measurements. By employing localized thermal excitation via focused laser beams with spot radii of 30-50 μm, our method circumvents the requirement for large-area heat flux. This not only enables sub-millimeter spatial resolution detection but also mitigates issues associated with large-area heat flux, such as exacerbated heat loss and non-uniform heat flux caused by wall heat conduction. Additionally, the method incorporates a three-dimensional thermal model to simulate heat diffusion in the multilayered wall, avoiding the assumption of one-dimensional heat conduction and thereby enhancing the accuracy of $h_c$ measurements. The smaller temperature excursion induced in the wall (~10 K) prevents significant temperature gradients in the fluid, further minimizing measurement errors. With limited sources of uncertainty, including propagated errors from input parameters and signal noise, the typical uncertainty of $h_c$ measured by the SPS method is approximately 7%. Overall, this method demonstrates high reliability, precision, and stability.

## 4. Conclusion

This study introduces an innovative approach for measuring the local intrinsic convective heat transfer coefficient—the Square-Pulsed Source (SPS) method. By using square-wave-modulated laser heating on the convective wall surface and another laser to detect wall temperature changes, the SPS method offers high efficiency, precision, and flexibility. It achieves a spatial resolution of less than 0.1 mm (spot diameter) and can accurately measure local convective heat transfer coefficients over 100 $W/(m^2 \cdot K)$ with a measurement error of less than 10%. In the air impinging jet experiments, the SPS method demonstrated excellent measurement performance, significantly enhancing both the accuracy and the measurable range of local convective heat transfer coefficients. This advancement represents a new tool for measuring local convective heat transfer, aiding in the further understanding and optimization of convective heat transfer processes in various applications.

## DATA AVAILABILITY

The data that support the findings of this study are available from the corresponding author upon reasonable request.

## DECLARATION OF COMPETING INTEREST

The authors have no known competing interest to declare.

## ACKNOWLEDGMENT

P.J. acknowledges support from the National Natural Science Foundation of China (NSFC) through Grant No. 52376058.

## REFERENCES

[1] D.-Y. Lee, K. Vafai, Comparative analysis of jet impingement and microchannel cooling for high heat flux applications, Int. J. Heat Mass Transfer, 42 (1999) 1555-1568.

[2] J. Town, D. Straub, J. Black, K.A. Thole, T.I. Shih, State-of-the-art cooling technology for a turbine rotor blade, Journal of Turbomachinery, 140 (2018) 071007.

[3] X. Li, J.F. Lovell, J. Yoon, X. Chen, Clinical development and potential of photothermal and photodynamic therapies for cancer, Nature reviews Clinical oncology, 17 (2020) 657-674.
[4] Y. Li, W. Li, T. Han, X. Zheng, J. Li, B. Li, S. Fan, C.-w. Qiu, Transforming heat transfer with thermal metamaterials and devices, Nature Reviews Materials, 6 (2020) 488 - 507.
[5] T.A. Moreira, A.R.A. Colmanetti, C.B. Tibiriçá, Heat transfer coefficient: a review of measurement techniques, Journal of the Brazilian Society of Mechanical Sciences and Engineering, 41 (2019) 264.
[6] Ş.-M. Simionescu, Ü. Düzel, C. Esposito, Z. Ilich, C. Bălan, Heat transfer coefficient measurements using infrared thermography technique, 2015 9th International Symposium on Advanced Topics in Electrical Engineering (ATEE), IEEE, 2015, pp. 591-596.
[7] L. Pagliarini, L. Cattani, M. Slobodeniuk, V. Ayel, C. Romestant, F. Bozzoli, S. Rainieri, Novel infrared approach for the evaluation of thermofluidic interactions in a metallic flat-plate pulsating heat pipe, Applied Sciences, 12 (2022) 11682.
[8] S. Han, R.J. Goldstein, The heat/mass transfer analogy for a simulated turbine endwall, Int. J. Heat Mass Transfer, 51 (2008) 3227-3244.
[9] M.-S. Chae, D.-Y. Lee, B.-J. Chung, Experimental study on local variation of buoyancy-aided mixed convection heat transfer in a vertical pipe using a mass transfer method, Exp. Therm Fluid Sci., 104 (2019) 105-115.
[10] D.N.N. Duarte, DIRECT TEMPERATURE GRADIENT MEASUREMENT USING INTERFEROMETRY, Experimental Heat Transfer, 12 (1999) 279-294.
[11] S. Bahl, J.A. Liburdy, Measurement of local convective heat transfer coefficients using three-dimensional interferometry, Int. J. Heat Mass Transfer, 34 (1991) 949-960.
[12] S.A. Putnam, S.B. Fairchild, A.A. Arends, A.M. Urbas, All-optical beam deflection method for simultaneous thermal conductivity and thermo-optic coefficient ( dn/dT) measurements, J. Appl. Phys., 119 (2016) 173102.
[13] M. Mehrvand, S.A. Putnam, Probing the Local Heat Transfer Coefficient of Water-Cooled Microchannels Using Time-Domain Thermoreflectance, J. Heat Transfer, 139 (2017) 112403.
[14] R.J. Murdock, S.A. Putnam, S. Das, A. Gupta, E.D. Chase, S. Seal, High-Throughput, Protein-Targeted Biomolecular Detection Using Frequency-Domain Faraday Rotation Spectroscopy, Small, 13 (2017).
[15] T. Germain, T.A. Chowdhury, J. Carter, S.A. Putnam, Measuring Heat Transfer Coefficients for Microchannel Jet Impingement Using Time-domain Thermoreflectance, 2018 17th IEEE Intersociety Conference on Thermal and Thermomechanical Phenomena in Electronic Systems (ITherm), 2018, pp. 449-454.
[16] M. Mehrvand, S.A. Putnam, Transient and local two-phase heat transport at macro-scales to nano-scales, Communications Physics, 1 (2018) 21.
[17] T. Chen, S. Song, Y. Shen, K. Zhang, P. Jiang, Simultaneous measurement of thermal conductivity and heat capacity across diverse materials using the square-pulsed source (SPS) technique, Int. Commun. Heat Mass Transf., 158 (2024) 107849.
[18] T. Chen, S. Song, R. Hu, P. Jiang, Comprehensive measurement of three-dimensional thermal conductivity tensor using a beam-offset square-pulsed source (BO-SPS) approach, Int J Therm Sci, 207 (2025) 109347.
[19] S. Sandell, E. Chávez-Ángel, A. El Sachat, J. He, C.M. Sotomayor Torres, J. Maire, Thermoreflectance techniques and Raman thermometry for thermal property characterization of nanostructures, Journal of Applied Physics, 128 (2020).

[20] T.L. Bergman, Fundamentals of heat and mass transfer, John Wiley & Sons2011.
[21] A. Saltelli, Global Sensitivity Analysis: the Primer, John Wiley & Sons2008.
[22] J. Yang, E. Ziade, A.J. Schmidt, Uncertainty analysis of thermoreflectance measurements, Rev. Sci. Instrum., 87 (2016) 014901.
[23] Y. Pang, P. Jiang, R. Yang, Machine learning-based data processing technique for time-domain thermoreflectance (TDTR) measurements, J. Appl. Phys., 130 (2021) 084901.
[24] P. Jiang, X. Qian, X. Gu, R. Yang, Probing Anisotropic Thermal Conductivity of Transition Metal Dichalcogenides MX2 (M = Mo, W and X = S, Se) using Time-Domain Thermoreflectance, Adv. Mater., 29 (2017) 1701068.
[25] P. Jiang, X. Qian, R. Yang, L. Lindsay, Anisotropic thermal transport in bulk hexagonal boron nitride, Physical Review Materials, 2 (2018) 064005.
[26] P. Jiang, D. Wang, Z. Xiang, R. Yang, H. Ban, A new spatial-domain thermoreflectance method to measure a broad range of anisotropic in-plane thermal conductivity, Int. J. Heat Mass Transfer, 191 (2022) 122849.
[27] T. Chen, S. Song, Y. Shen, K. Zhang, P. Jiang, Simultaneous measurement of thermal conductivity and heat capacity across diverse materials using the square-pulsed source (SPS) technique, International Communications in Heat and Mass Transfer, 158 (2024) 107849.
[28] L. Huang, M.S. El-Genk, Heat transfer of an impinging jet on a flat surface, Int. J. Heat Mass Transfer, 37 (1994) 1915-1923.
[29] M.J. Assael, K.D. Antoniadis, J. Wu, New Measurements of the Thermal Conductivity of PMMA, BK7, and Pyrex 7740 up to 450K, Int. J. Thermophys., 29 (2008) 1257-1266.
[30] Y. Touloukian, E. Buyco, Thermophysical Properties of Matter-The TPRC Data Series. Volume 4. Specific Heat-Metallic Elements and Alloys, DTIC Document, 1971.
[31] D.R. Lide, CRC handbook of chemistry and physics, CRC press2004.
[32] H. Martin, Heat and mass transfer between impinging gas jets and solid surfaces, Advances in heat transfer, Elsevier1977, pp. 1-60.